# Cryogenically Enhanced Laser-Induced Amorphous Phase Transitions in Crystalline Silicon

Conrad Kuz[1,*], Andy Lee[1], Shashu Tomar[1], Ravleen Kaur[2], Mohamed Yaseen Noor[2], Justin Twardowski[2], Liam Clink[1], Roberto C. Myers[1,2,3], Enam Chowdhury[2,1,3,†]
[1]Department of Physics, The Ohio State University, Columbus, OH 43228, USA
[2]Department of Materials Science and Engineering, The Ohio State University, Columbus, OH 43228, USA
[3]Department of Electrical & Computer Engineering, The Ohio State University, Columbus, OH 43228, USA

**ABSTRACT**. Amorphization of silicon is crucial to applications in photonics, microelectronics and solar cell technologies. Ultrafast lasers have been used to generate amorphous silicon from crystalline silicon using rapid non-thermal melting and solidification in room temperature. As material temperature can affect cooling rates significantly, adding temperature control in ultrafast laser modification of silicon may allow a new degree of freedom in ultrafast laser modification. In this work, we investigate the role of cryogenic temperature in governing ultrafast damage pathways via single-shot femtosecond laser irradiation of silicon from room temperature down to 24 K at 1030 nm. Across this temperature range, we observe a pronounced enhancement of amorphization at lower temperatures, revealed through optical microscopy, Raman spectroscopy, and Kelvin probe force microscopy (KPFM). Raman analysis identifies this ring as an amorphous surface layer, while complementary AFM and SEM imaging show temperature-dependent changes in surface morphology, including localized melt redistribution and refrozen material. To elucidate the physical origins of this behavior, we implement a carrier dependent two-temperature model (nTTM). The simulations reproduce the experimentally observed trends and indicate that reduced phonon population, modified absorption pathways, and altered lattice relaxation dynamics at cryogenic temperatures collectively promote amorphous freezing over recrystallization. This study represents the first detailed examination of silicon under ultrafast irradiation below the liquid-nitrogen regime and reveals temperature-governed mechanisms relevant for advanced silicon microstructuring.

## I. INTRODUCTION.

Amorphous silicon (a-Si) has become an established material for functionality in solar cell technologies [1,2] and photonic waveguides [3]. Ultrafast pulsed laser irradiation enables rapid non-thermal melting and solidification, allowing controlled transitions between crystalline and amorphous phases [4,5]. Laser-induced amorphization of silicon has gained interest for its applications in waveguide writing [3,6], solar cell engineering [5,7], and selective etching [8]. Despite this broad technological potential, fundamental questions remain regarding how silicon responds to ultrafast laser excitation under different thermodynamic conditions, especially at cryogenic temperatures, where phonon populations and transport properties deviate significantly from room-temperature characteristics.

While the influence of temperature on silicon laser processing has been documented as far back as 1988 [9], low temperature effects on the initiation of laser damage remains an under-explored field. Studies have observed shifts in laser-induced damage thresholds (LIDT) [10–13] and surface patterning [14–17] changes with different temperatures, yet almost no research has investigated these phenomena below liquid nitrogen temperatures. To the best of our knowledge, this work represents the first systematic investigation of ultrafast laser-induced amorphization in silicon at temperatures as low as 24 K. Transitioning to this extreme cryogenic regime fundamentally alters the material's thermodynamic response. A decrease in phonon population at low temperatures limits absorption for 1030 nm pulses, which rely on phonon-assisted transitions. Additionally, the subsequent relaxation and solidification dynamics are dominated by increases in heat transport. Specifically, silicon's thermal conductivity increases significantly at cryogenic temperatures [18], leading to faster thermal diffusion. This enhanced cooling rate plays a decisive role later in the process, as the competition between rapid thermal dissipation and solidification fronts ultimately determines whether the lattice recrystallizes or freezes into the observed amorphous state.

The interaction of ultrafast laser pulses with crystalline silicon is governed by a complex interplay of carrier excitation mechanisms. For a 1030 nm pulse at room temperature, the photon energy slightly exceeds silicon's indirect bandgap. However, linear absorption is fundamentally limited by silicon's indirect band structure, which requires phonon scattering to conserve momentum [19]. Consequently, under high-intensity femtosecond irradiation, carrier generation is driven strongly by nonlinear processes, specifically two-photon absorption (TPA) [20]. This rapid energy deposition creates a dense electron-hole plasma, driving the system into a highly non-equilibrium state where the free electron temperature spikes significantly while the lattice initially remains cold. Accurately simulating this highly non-equilibrium state requires a numerical framework that simultaneously tracks the evolution of the lattice temperature $T_l$, the electronic temperature $T_e$ and the free carrier density $n_e$. This is typically achieved using a carrier-dependent two-temperature model (nTTM), which accounts for the dynamic coupling between initial energy deposition, carrier-driven instabilities, and lattice thermalization [21].

*Contact author: kuz.1@osu.edu
†Contact author: chowdhury.24@osu.edu

Following this initial energy deposition, the phase transition mechanism is dictated by the peak carrier density. In the standard thermal regime, hot carriers transfer energy to the lattice via electron-phonon coupling, raising the lattice temperature until it exceeds the melting point. However, if the carrier density is increased, the material undergoes "non-thermal melting." In this regime, the high concentration of excited electrons modifies the interatomic potential energy surface, inducing lattice instability and disorder on a sub-picosecond timescale, faster than the thermal diffusion and electron-phonon coupling [22]. Similarly, there are two pathways to ablation, the carrier density can reach an even higher critical point (typically $\sim 10^{22}\text{cm}^{-3}$), causing the lattice to lose its structural integrity, or the lattice can reach a critical temperature causing boiling and ejection of material [23]. The final structural state, whether the material recrystallizes or freezes into an amorphous phase, is then determined by the cooling rate and the velocity of the solidification fronts [24].

At near threshold fluences, the entire irradiated volume undergoes rapid melting and quenching, creating a fully amorphous spot. As fluence increases, the center of the melt pool retains heat longer, allowing for epitaxial regrowth through slower cooling [25]. This results in a distinctive amorphous ring surrounding a recrystallized center, a morphology defined by the competition between rapid interfacial cooling at the periphery and sustained thermal energy at the focus. In silicon, this crystalline regrowth rate is strongly orientation dependent. Si (100), studied here, recrystallizes faster compared to other crystal orientations such as (110) or (111) [26–28]. Due to the faster supported recrystallization of Si (100), it is more resistant to amorphization from laser irradiation [29]. This orientation-dependent resistance defines the standard behavior at room temperature, yet the impact of extreme cold, where thermal diffusivity and carrier dynamics change drastically, on this competition remains unexplored.

This work seeks to determine how silicon responds to ultrafast laser pulses across the cryogenic regime, with a particular focus on identifying the fundamental mechanisms that drive temperature-dependent transitions between crystalline, molten, and amorphous states. By extending observations down to 24 K, we aim to reveal how decreased absorption and altered lattice relaxation dynamics promote amorphous freezing over recrystallization. Through a combination of surface characterization methods and carrier-dependent two-temperature model (nTTM) simulations, this study provides a comprehensive framework for understanding laser-induced phase transitions under extreme thermodynamic conditions.

## II. METHODS

### A. Experimental

Single-crystal (100) silicon samples were cleaved from a larger wafer and ultrasonically cleaned in a 1:1 mix of methanol and acetone. To remove the native oxide, samples were etched in a 5% hydrofluoric acid (HF) solution for 10 minutes. The resulting hydrogen-terminated surface was verified with a change in water contact angle. The sample was transferred to vacuum in the cryostat within 15 minutes to prevent re-oxidation.

Cryogenic cooling of the silicon was provided by an ARS DMX-20 system. The cryostat keeps the samples in a vacuum of approximately $5\text{x}10^{-3}$ mbar. Temperature is controlled from 300 K down to 24 K via a resistive heater and PID loop. A custom stage was designed to move the entire cryostat, with ~μm precision. This allows adequate control to move the sample for laser irradiation and to ensure the sample is in the focus.

Irradiation was performed using a commercial PHAROS laser system (300 fs, 1030 nm, 200 μJ). The beam was focused through the cryostat window using a 75 mm lens to a 19.2 μm FWHM spot. A half-waveplate and polarizer controlled the pulse energy, which was calibrated to account for transmission loss through the cryostat window. A 10x objective was used for focal spot imaging with the sample removed, while a 10x objective at a 20° angle of incidence provided real-time surface monitoring during damage. The setup is shown in the Supplementary Material [36].

The cryostat was set to evenly spaced temperatures, and samples were held for equilibration for 10 minutes before irradiation. At each temperature, the same fluences were tested. Fluences are chosen to be above the damage threshold of ~0.2 $\text{Jcm}^{-2}$ [30–32] for silicon. Irradiation was performed at fluences ranging from 0.5 to 1.75 $\text{Jcm}^{-2}$ by increments of 0.25 $\text{Jcm}^{-2}$. Single shot irradiation is performed as this allows for better study of fundamental mechanisms, instead of compounding effects of multi-shot irradiation. For each fluence and temperature combination, 8 damage spots were made for statistical verification.

Following laser exposure, the samples were characterized using a suite of complementary techniques to capture both surface morphology, electronic property changes, and phase change. Optical microscopy was employed first to document overall damage patterns and damage size. Atomic Force Microscopy (AFM), using a Nanosurf Flex, provided topographical measurements, enabling nm scale height and roughness analysis. KPFM was used to map local surface potential variations, providing insight into changes in electronic structure. Scanning Electron Microscopy (SEM), using a Thermo Scientific Apreo, provided imaging of microstructural features such as cracks, voids, and bubble formations, which are critical for understanding temperature-dependent damage mechanisms. Finally, Raman spectroscopy using Renishaw Invia, with a 514 nm excitation, was used to characterize the extent of amorphization and recrystallization. The collected Raman spectra were processed using RamanSPy [33] and subsequently fitted in Python.

**B. Simulation**

A 2-dimensional nTTM is used to simulate behavior of the lattice and electrons under irradiation. The model is implemented in a 2D cylindrical (r–z) geometry, which captures both radial and depth-dependent heat diffusion. This axisymmetric configuration reflects the Gaussian spatial profile of the laser beam and enables simulation of multidirectional thermal transport during and after the pulse.

$$\frac{\partial n_e}{\partial t} = \frac{\alpha_{SPA} I(r,z,t)}{\hbar\omega} + \frac{\beta I^2(r,z,t)}{2\hbar\omega} - \gamma n_e^3 + \theta n_e + \nabla(D_0 \nabla n_e) \tag{1}$$

$$C_e \frac{\partial T_e}{\partial t} = \nabla\cdot(K_e \nabla T_e) - G(T_e - T_l) + S(r,z,t) - \frac{\partial n_e}{\partial t}\left(E_g + 3K_B T_e\right) - n_e\left(\frac{\partial E_g}{\partial n_e}\frac{\partial n_e}{\partial t} + \frac{\partial E_g}{\partial T_l}\frac{\partial T_l}{\partial t}\right) \tag{2}$$

$$C_l \frac{\partial T_l}{\partial t} = \nabla\cdot(K_l \nabla T_l) + G(T_e - T_l) \tag{3}$$

Eq. (1) describes the evolution of the free carrier density $n_e$ over time $t$. The first two terms on the right-hand side account for carrier generation through linear absorption $\alpha_{SPA}$ and two-photon absorption $\beta$, with corresponding photon energies $\hbar\omega$ and $2\hbar\omega$. The third term represents the loss of carriers via Auger recombination at a rate determined by $\gamma$. Impact ionization is accounted for by the fourth term using the coefficient $\theta$, which describes avalanche carrier generation. Finally, the last term characterizes the spatial redistribution of the plasma through ambipolar diffusion, governed by the coefficient $D_0$.

The evolution of the electronic temperature $T_e$ is governed by Eq. (2), which describes the energy balance within the carrier system. Electronic heat capacity is given by $C_e$. The first three terms on the right-hand side represent electronic thermal conduction with conductivity $K_e$, energy transfer to the lattice via the electron-phonon coupling constant $G$, and the laser source term $S(r,z,t)$, respectively. The fourth term accounts for the energy consumed or released during carrier generation and recombination, where $E_g$ is the bandgap energy and $3K_B T_e$ represents the average kinetic energy of the carriers. The final bracketed term describes bandgap renormalization, capturing the shift in $E_g$ as a function of both the carrier density $n_e$ and the lattice temperature $T_l$.

The evolution of the lattice temperature $T_l$ is described by Eq. (3), representing the thermal balance of the silicon crystal structure. Lattice heat capacity is given by $C_l$. The first term on the right-hand side accounts for lattice thermal conduction with conductivity $K_l$. The second term represents the energy gain from the electronic system through the electron-phonon coupling constant $G$. This coupling term ensures that the energy initially deposited into the carrier plasma is transferred to the lattice, eventually driving the structural changes observed in the experiment.

The spatial and temporal distribution of the laser intensity $I(r,z,t)$ as it propagates through the silicon is described by Eq. (4). This formulation accounts for linear attenuation due to both single-photon absorption $\alpha_{SPA}$ and free-carrier absorption $\alpha_{FCA}$, as well as nonlinear losses from two-photon absorption $\beta$. The local energy deposition rate is then defined by the laser heat source term $S(r,z,t)$ in Eq. (5). This term aggregates all contributing absorption pathways to determine the rate at which energy is coupled into the electronic system.

$$\frac{\partial I(r,z,t)}{\partial z} = -(\alpha_{SPA} + \alpha_{FCA})I - \beta I^2 \tag{4}$$

$$S(r,z,t) = (\alpha_{SPA} + \alpha_{FCA})I(r,z,t) + \beta I(r,z,t)^2 \tag{5}$$

Intensity at the top layer (z=0) is modeled as a Gaussian in time and space of corresponding peak fluence with carrier and temperature dependent surface reflectivity considered. The simulation starts at $t = 0$ with the peak of the Gaussian pulse at $t = 600$ fs.

To adapt the numerical framework for extreme cryogenic conditions, extending beyond the temperature ranges typically modeled in existing literature, several key parameters were modified to reflect the physical transitions. Specifically, the linear absorption coefficient $\alpha_{SPA}$ was extracted from experimental measurements [34] to account for the significant reduction in photon absorption at low temperatures where the phonon population "freezes out," reducing the rate of indirect, phonon-assisted transitions. Furthermore, the lattice heat capacity $C_l$ [35] and thermal conductivity $K_l$ [18] were modeled using cryogenic-specific data to capture the rapid increase in thermal diffusivity characteristic of the extreme cold regime.

The simulation size is a radius of $r = 25$ μm and a maximum depth of $z = 5$ μm. A minimum grid spacing of 2 nm in $z$ and 100 nm in $r$ is used. Neumann boundary conditions are used on all edges. Dynamic time steps are calculated using the Courant-Friedrichs-Lewy (CFL) condition. For additional detail on the simulation, please refer to the Supplementary Material [36].

## III. RESULTS AND DISCUSSION

### A. Experimental

Experimental irradiation of silicon from 300 K down to 24 K reveals a pronounced shift in surface morphology characterized under optical microscopy by the emergence of a distinct, annular feature. The ring is attributed to an amorphous zone, as evidenced by higher reflectivity in the visible range compared to crystalline silicon [36], and later verified by Raman spectroscopy. At room temperature, Si (100) typically exhibits high resistance to amorphization [29] due to its high critical solidification velocity; however, as the initial temperature decreases, this amorphous ring becomes increasingly prominent. At 24 K, the stability of the amorphous ring suggests that the cryogenic environment increases the quench rate above the epitaxial crystalline regrowth limit of Si (100). Consequently, the (100) surface exhibits an amorphization response similar to the more susceptible (111) orientation. Optical microscopy confirms that while the ring is faint or absent at 300 K, it appears consistently across all tested fluences at cryogenic temperatures, with its diameter and average intensity increasing as the sample is cooled.

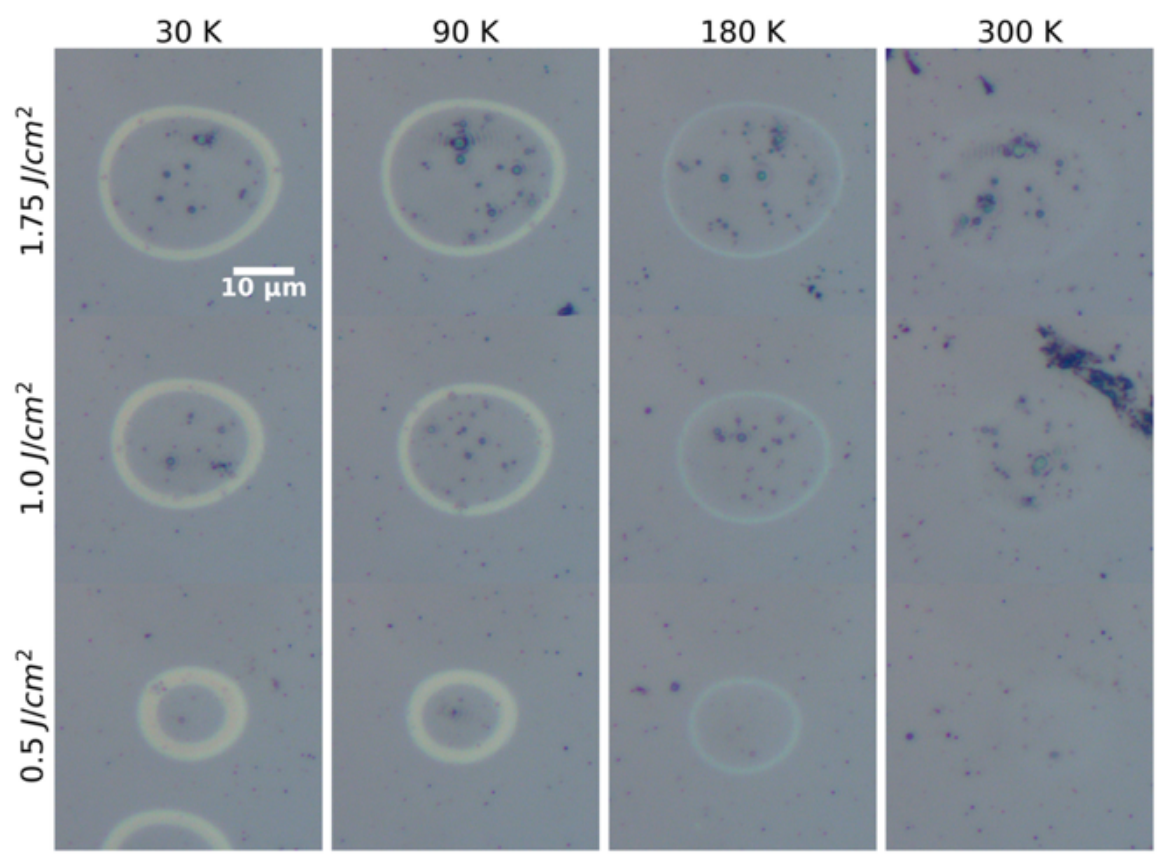

FIG. 1. Optical microscopy of single-shot damages at various temperatures and fluences. The amorphous ring size is shown to grow with decreasing temperature. A full grid of all temperatures and fluences is available in the Supplementary Material [36].

The evolution of these surface features is illustrated in Fig. 1, which displays the damage morphology across a matrix of initial temperatures and laser fluences. At 300 K, the irradiation sites exhibit a central damage crater characterized by localized nanoscale pits, indicative of breakdown at pre-existing surface defects, and a narrow, almost indistinct amorphous periphery. As the sample is cooled to 180 K and 90 K, a well-defined bright annular ring emerges, which systematically broadens as the temperature decreases. By 30 K, the ring structure is a strong feature of the damage site for all fluences, exhibiting a significant increase in both radial extent and optical brightness. Notably, while the ring diameter scales with the incident fluence, consistent with the expanding area that exceeds the amorphization threshold, the contrast of the ring remains consistently higher at cryogenic temperatures. This suggests that the extreme cold not only promotes the initiation of the amorphous phase but also leads to a more robust or thicker amorphous layer compared to room-temperature irradiation.

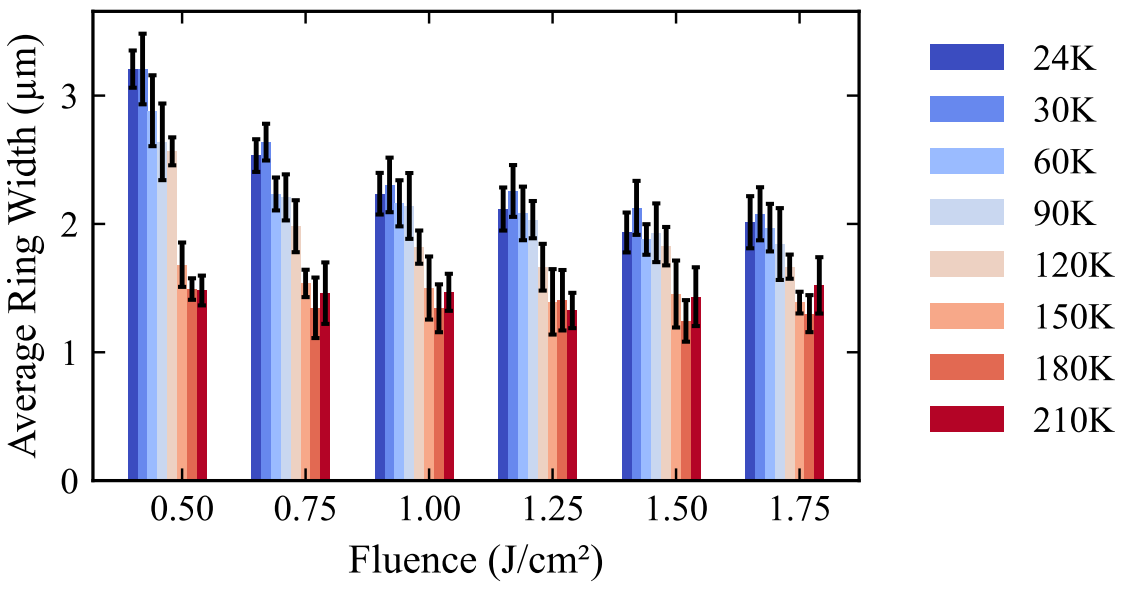

FIG. 2. Analysis of amorphous ring widths from optical microscopy compared to fluence and temperature. Lower temperature and fluence increases the range of amorphization

To quantify the temperature-dependent evolution of these features, the radial width of the amorphous ring was measured across the full range of incident fluences and initial temperatures. This metric, representing the spatial extent of the region where rapid melting is followed by amorphous freezing, is plotted in Fig. 2. The measurement reveals a trend where the ring width expands as the initial temperature is lowered, reaching its maximum extent at 24 K. This widening confirms that at cryogenic temperatures, the steep thermal gradient between the central focal spot and the surrounding lattice promotes a broader zone where the solidification velocity exceeds the recrystallization threshold velocity. Notably, this temperature-dependent widening is more pronounced at lower fluences. The radial expansion of the amorphous ring exhibits greater sensitivity to initial temperature at lower incident fluences. This indicates that near the amorphization threshold, reductions in lattice temperature significantly enhance the radial extent of increased quench rate required to stabilize the amorphous phase. Data points above 210 K were excluded from quantitative analysis in Fig. 2, as the resulting amorphous feature widths were too faint to permit reliable measurement.

Applying the Liu ($D^2$) method [37] to the inside and outside of the amorphous rings reveals a distinct splitting in the damage thresholds at cryogenic temperatures. As shown in Fig. 3, the extrapolated damage thresholds for the outer ring (onset of amorphization) and the inner ring (central recrystallization/ablation) exhibit markedly different temperature sensitivities. Interestingly, extrapolating the linear fits of the cryogenic data suggests an intersection point near 300 K. This convergence implies that at room temperature, the "window" for forming a distinct amorphous ring effectively vanishes, as the fluence required to initiate amorphization becomes indistinguishable from the threshold for central damage or ablation. The damage thresholds measured above 210 K (green points) fall between these extrapolated lines, consistent with a regime where rapid recrystallization destabilizes the annular structure, preventing the formation of the discrete "inner" and "outer" phases observed at low temperatures.

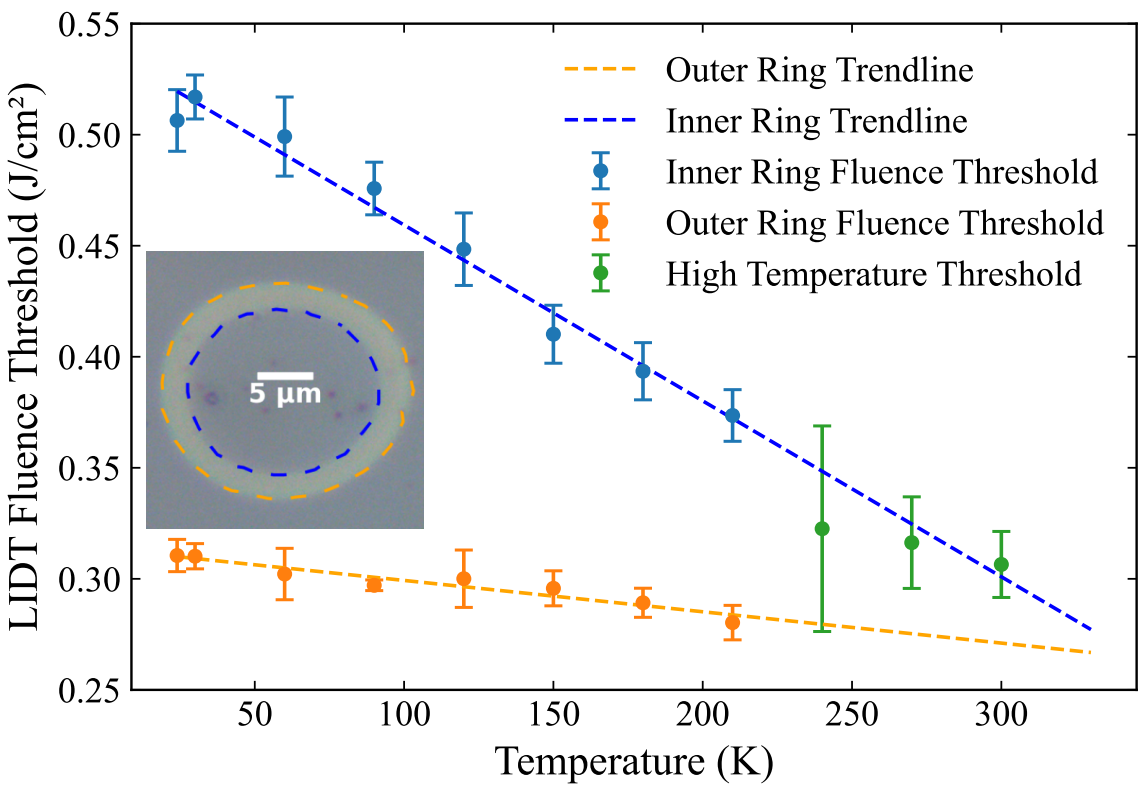

FIG. 3. LIDT of the inner and outer rings representing the amorphous zone. The high temperature points, where measurement of the bound is indistinct, fall between the predicted outer and inner ring trendlines. The inset image demonstrates the differentiation between the inner and outer rings under optical microscopy.

The distinct slopes in Fig. 3 imply that different physical mechanisms govern the inner and outer damage thresholds. The inner ring threshold, associated with central ablation or permanent structural damage, shows a marked dependence on the initial temperature. This sensitivity may reflect the interplay of several temperature-dependent parameters, such as increased thermal and carrier diffusion rates ($K_l, D_{amb}$) or the reduction in linear absorption ($\alpha_{SPA}$) as phonon populations freeze out. In contrast, the outer ring threshold remains relatively stable across the tested range. This behavior is consistent with a melting process dominated by the latent heat of fusion, where the latent heat energy is large compared to the small energy differences between varying initial temperatures. The divergence of these thresholds at lower temperatures effectively opens a thermodynamic process window where amorphization is favored. As the temperature approaches 300 K, this window narrows, suggesting that high-temperature damage results from a complex competition between ablation and recrystallization absent of the formation of a distinct amorphous phase.

SEM imaging (Fig. 4(a)) reveals a distinct, dark annular region corresponding to the amorphous phase due to the difference in secondary electron emission, exhibiting high homogeneity with minimal debris or redeposition. While the primary ring structure is uniform, isolated nanoscale pits are observed scattered in the periphery. Since the distribution of these features varies randomly between irradiation sites, they are attributed to localized breakdown at pre-existing surface imperfections or the redeposition of debris ejected from the central ablation crater. Crucially, the absence of periodic ripples (e.g. LIPSS) or chaotic resolidified melt within the main annular zone suggests a non-ablative melting and rapid solidification process, distinct from the ablation observed in the central crater. The evolution of surface morphology across the full range of fluences and temperatures is documented in the Supplementary Material [36]. Notably, while the outer amorphous ring retains its structural homogeneity across the tested range, the central impact zone, where the local fluence exceeds the ablation threshold, exhibits characteristic laser-induced periodic surface structures (LIPSS) and resolidified melt artifacts. This spatial segregation confirms that the gentle, topographically flush phase transformation is confined to the peripheral thermodynamic window, distinct from the carrier-driven ablation mechanisms dominating the focal center.

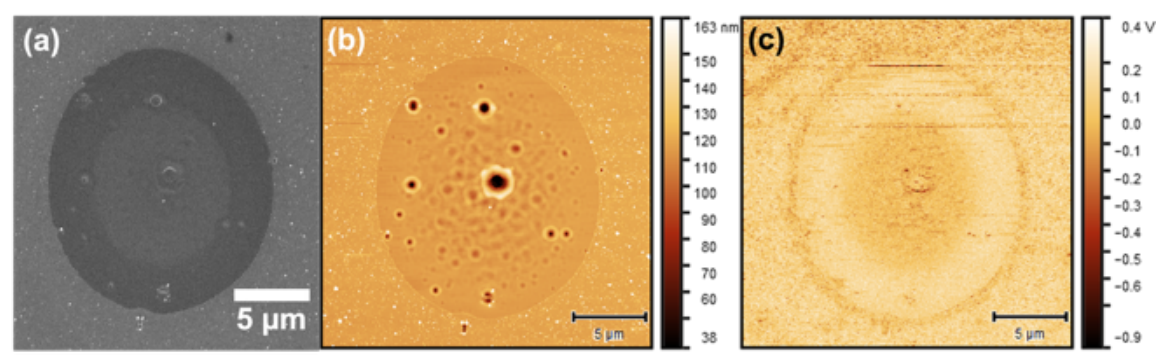

FIG. 4. The same laser damage spot produced at 0.5 Jcm$^{-2}$ at 24K observed under (a) SEM, (b) AFM, (c) KPFM.

AFM topography (Fig. 4(b)) further characterizes this modification, revealing that the annular region maintains a height profile slightly depressed ($\approx$ 4 nm) relative to the surrounding pristine substrate. This minor topological shift indicates that the phase transformation occurs with minimal mass transport, distinct from the cratering and roughness characteristic of the central ablation regime. The resulting surface remains quasi-planar, effectively creating a smooth, buried interface suitable for photonic applications.

KPFM, measuring the contact potential difference $V(CPD)$ $=1/e$ $(\phi_{\text{tip}}-\phi_{\text{sample}})$, (Fig. 4(c)) detects a pronounced, uniform shift in surface work function across the ring. The brighter annular region corresponds to a higher measured surface potential, indicating a reduced local work function relative to crystalline silicon. The CPD, relative to the unmodified silicon, of the amorphous zone is noted to increase by 0.14 V from 300 K to 24 K. See the Supplementary Material for further details of KPFM [36]. This reduction is consistent with laser-induced amorphization, which introduces dangling bonds and shifts the Fermi level position. The uniform bright contrast therefore confirms a stable amorphous phase with altered electronic structure rather than simple surface roughness or damage.

While KPFM provides a map of the electronic surface states, it offers only indirect evidence of the atomic arrangement. To definitively identify the structural composition of the annular features and quantify the extent of the phase transformation, spatially resolved Raman spectroscopy was performed on the laser-modified regions. Two peaks are of importance, a sharp peak at $520\ \text{cm}^{-1}$ representing crystalline silicon, and a broad peak at centered at 480 cm$^{-1}$ representing the amorphous phase of silicon. The spectral maps in Fig. 5 display the intensity of the phonon mode distinctive of amorphous silicon ($a$-Si).

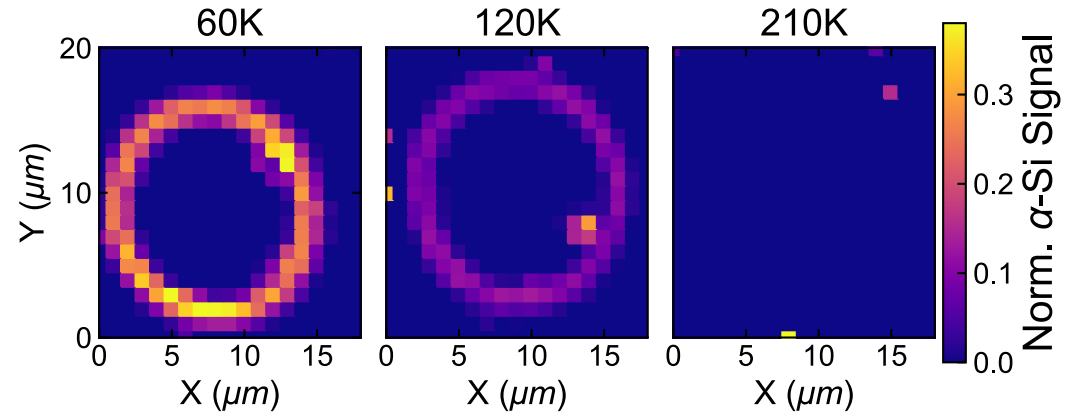

Fig. 5. Spatial maps of the amorphous Raman signal, calculated as the area of the amorphous peak normalized by the area of the crystalline peak, of silicon laser damage at 0.5 Jcm$^{-2}$ with initial temperatures of 60, 120, and 210 K. The amorphous amplitude at ~480 cm$^{-1}$ corresponds with the location of the rings under optical.

The Raman maps confirm that the bright annular features observed in optical microscopy correspond to a continuous region of amorphous silicon. At cryogenic temperatures (60 K and 120 K), the amorphous signal forms a robust, high-intensity ring that spatially correlates with the radial width optical measurements previously discussed. As the substrate temperature is increased to 210 K, the amorphous amplitude drops and becomes undetectable. This trend aligns with the optical data, confirming that above 210 K, the rate of recrystallization allows for full restoration of the crystalline lattice, preventing the amorphous phase.

By analyzing the intensity ratio between the integrated crystalline ($I_c$) phonon modes in the amorphous ring and background silicon, the effective thickness of the amorphous layer ($d_{a\text{-Si}}$) was calculated using the absorption-dependent scattering model [36]. Note that our measurements may be overestimated by a few nm, as described in the Supplementary Material [36], but still provides a good quantitative approximation. As shown in Fig. 6(a), the maximum amorphous layer thickness exhibits a strong inverse dependence on the initial substrate temperature. At 30 K, the layer reaches a maximum thickness of $\approx 35$ nm, indicating that the phase transformation is not merely a surface effect but extends noticeably into the bulk. This thickness approaches the 50 nm depths observed with mid-infrared (MIR) pulses [38], where field-driven absorption regime facilitates deeper energy deposition. As the temperature rises, this thickness decreases, effectively reaching zero above the 210 K threshold.

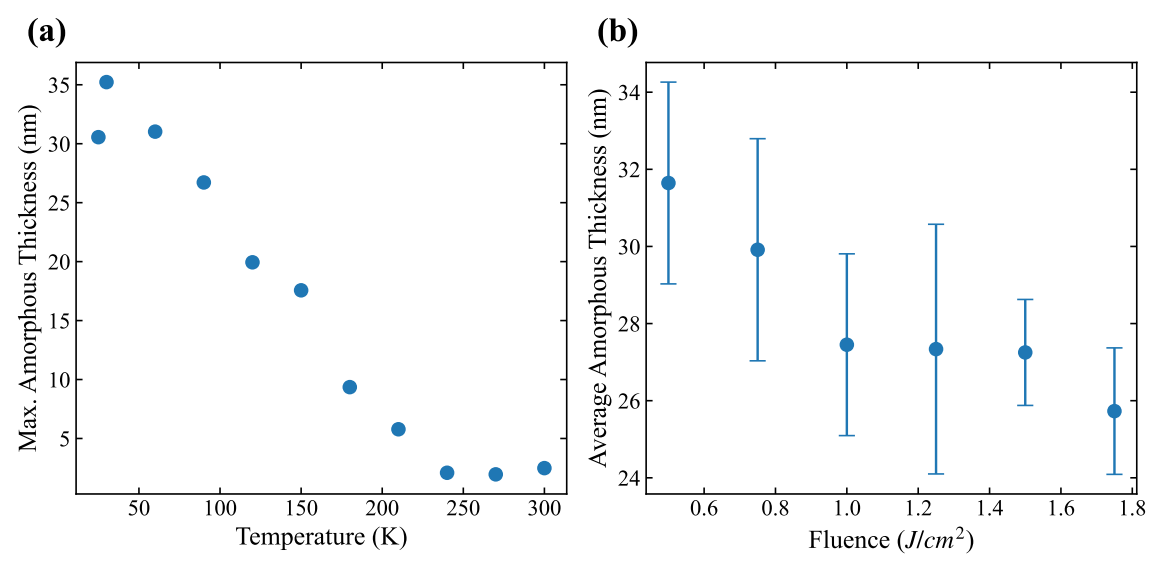

Fig. 6. (a) Maximum measured amorphous layer thickness compared to temperature measured by Raman for 0.75 J/cm^2. (b) Average amorphous layer thickness compared to fluence for 30 K.

The evolution of the amorphous layer thickness, calculated from the Raman intensity ratios, reveals a physically consistent competition between quench rates and crystal regrowth. As shown in Fig. 6(b), the average amorphous layer thickness decreases only slightly at higher fluences, even as the incident laser fluence is increased. This behavior suggests that the phase transformation is locked to a specific local fluence threshold, visualized as a "wing" of the Gaussian spatial profile where the energy density matches the critical amorphization value ($F_{th}$). Increasing the peak pulse energy merely pushes this threshold condition radially outward, without significantly increasing the local energy deposition at the ring's location. The slight reduction in thickness at higher fluences likely arises from lateral heat diffusion; as the peak fluence rises, the hotter central region (which undergoes ablation or rapid recrystallization) acts as a thermal reservoir. This excess heat slows the cooling rate at the peripheral ring, allowing the crystalline solidification front, propagating upwards from the bulk, more time to consume the molten silicon before it freezes into the amorphous state.

### B. Simulation

The two-temperature model simulation was run for 15 ns for two initial silicon temperatures, 300 K and 30 K, providing sufficient time to capture both ultrafast energy deposition and subsequent lattice cooling. These temperatures represent the extremes of the experimental range. Energy deposition in the model is quantified through the evolution of the free-carrier density and carrier temperature, which together determine how much absorbed energy is ultimately transferred into lattice heating. Cryogenic cooling alters both carrier generation and thermal transport, leading to significant differences in energy deposition between the two initial temperatures.

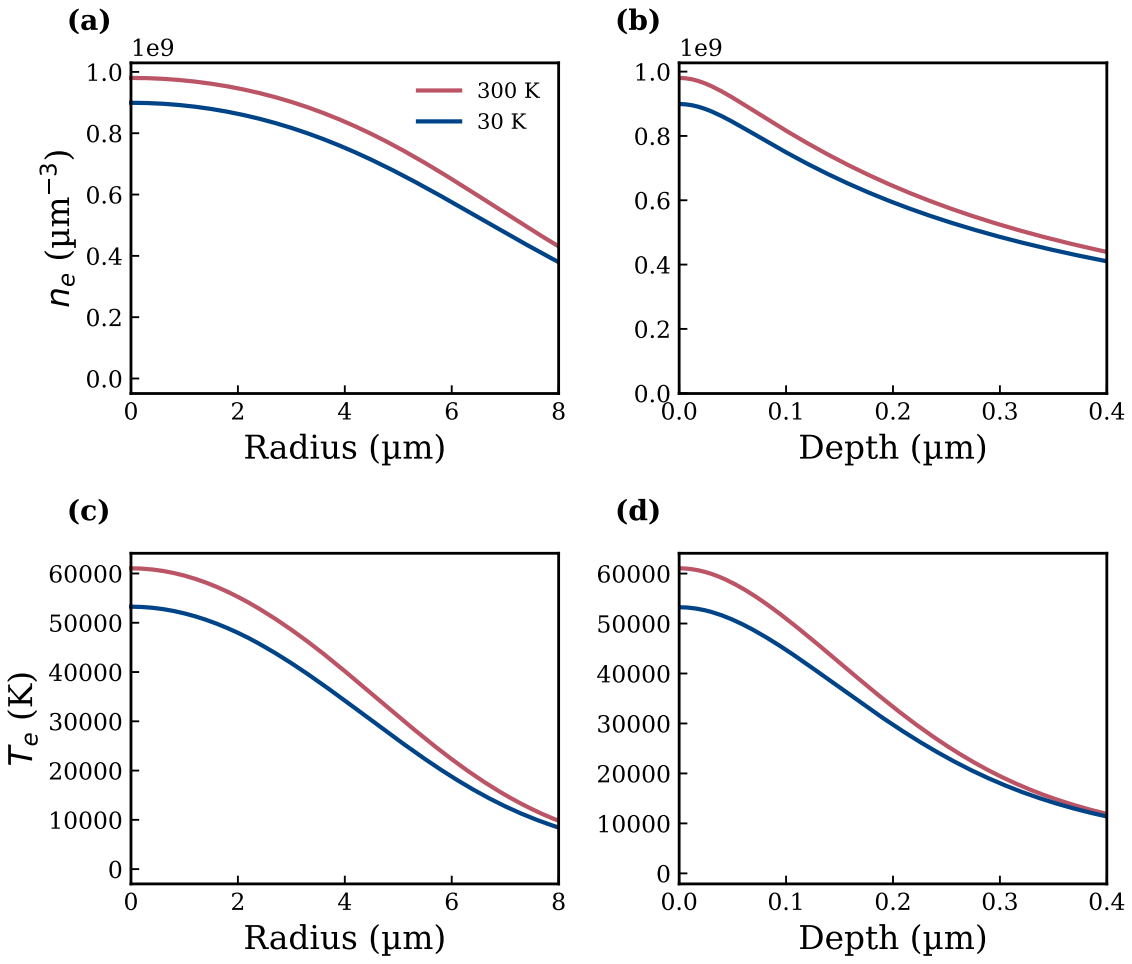

Fig. 7. Distribution of carrier density along the surface (a) at the top layer and depth (b) at the center and carrier temperature across surface (c) and depth (d) of the silicon at 1.5 ps. The largest differences occur at the top center portion.

Spatial profiles of electron temperature and carrier density are displayed in Fig. 7 at 1.5 ps. At this point in time, the pulse is effectively done depositing energy. The largest difference in energy deposition occurs at the spatial center of the pulse. The difference becomes less pronounced toward the edges. Reducing the substrate's initial temperature from 300 K to 30 K yields an 8% ($8 \times 10^7$ μm$^{-3}$) reduction in carrier density and a 13% ($8 \times 10^3$ K) decrease in carrier temperature. This temperature and carrier generation difference is significant when considering total energy deposition. The electronic heat capacity in the nTTM is defined as $C_e = 3 n_e k_B$, meaning that increasing the carrier density directly increases the heat capacity of the electron–hole plasma. The total electronic energy therefore scales with the product of carrier density and carrier temperature. The difference in energy is attributed mostly to absorption before the temporal peak of the pulse for two reasons. First, SPA of the material at 30 K, despite initially being two orders of magnitude smaller, quickly increases and matches the SPA of the 300 K material near the peak of the pulse. Second, the two-photon absorption has the largest impact of generation of carriers for most of the pulse. See the Supplementary Material for analysis of absorption dynamics and the competing magnitude of SPA and TPA during the pulse [36].

The reduced energy deposition facilitates more rapid thermal dissipation during the cooling phase by minimizing the total enthalpy within the irradiated volume. This lower thermal load, coupled with the enhanced thermal conductivity ($K_l$) characteristic of cryogenic regimes, establishes the steep temperature gradients necessary for amorphization. These dynamics directly contribute to the elevated quench rates analyzed as follows.

The nTTM implementation in this study employs an effective heat capacity method to account for the latent heat of fusion, which stabilizes the lattice temperature near the melting point ($T_m$) but limits the direct resolution of the solidification interface velocity. Despite this constraint, the model accurately captures the relative thermal transients and transport properties critical to the cryogenic regime. For this comparative study, the lattice quench rate, defined as the maximum $dT_l/dt$ following peak thermalization outside of the latent transition zone, is utilized as a physical proxy for the amorphization window. This approach is consistent with other studies which identify a critical quench rate on the order of 0.1 K/ps as the threshold for the kinetic arrest of epitaxial regrowth in silicon [24,39,40]. The simulation's quench rate along the surface and through the depth at the melted region is shown in Fig. 8.

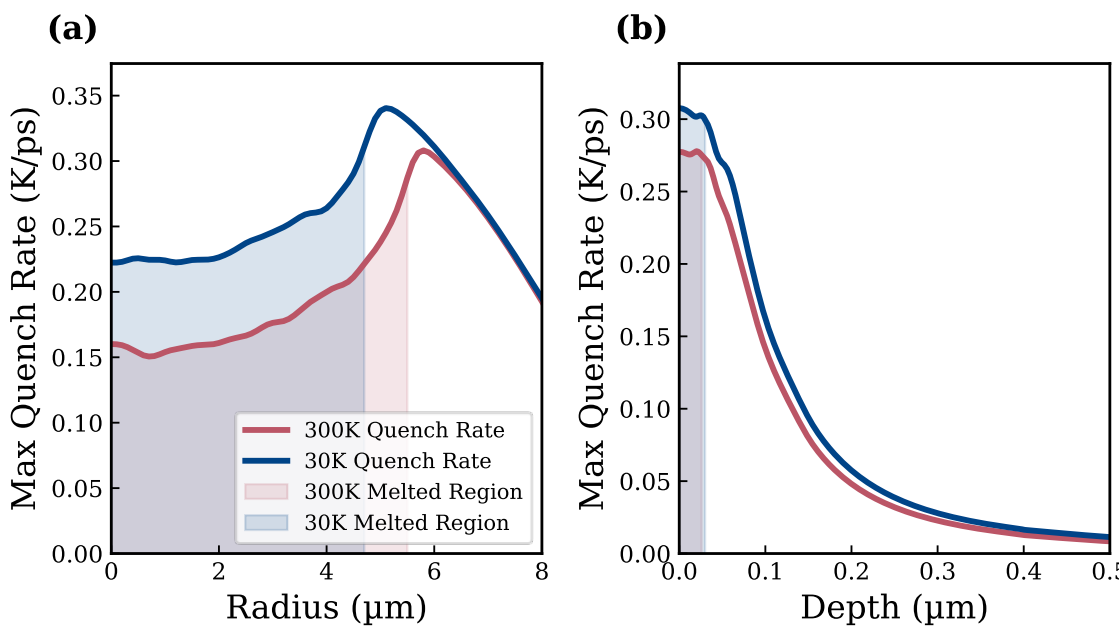

Fig. 8. Quenching rate of silicon at 0.5 Jcm$^{-2}$ across the surface. (b) Quenching rates along depth profile at the corresponding widest melt radius of r=4.7 μm at 30 K and at 300 K r=5.5μm. The 30 K substrate demonstrates a much higher quench rate.

The simulated quench rates along the surface, shown in Fig. 8(a), exhibit a marked divergence between the cryogenic and room-temperature regimes, with the 30 K substrate supporting an average cooling rate approximately 40% higher on average than the 300 K baseline within the melt zone. In both cases, the quench rate peaks near the melt-zone boundary, which correlates with the spatial localization of the amorphous ring observed in the experiment. While the 300 K simulation shows a sharp localized increase in quenching sufficient to produce the faint annular features seen in optical microscopy (Fig. 1), the 30 K substrate sustains elevated quench rates across a significantly broader radial extent. Thermodynamically, this intensified quenching generates the steep temperature gradients required to drive the solidification interface velocity beyond the critical threshold for Si (100) recrystallization.

At the periphery, the simulated melt depth (~ 30 nm) for cryogenic conditions aligns closely with the amorphous layer thickness derived from Raman intensity ratios (Fig. 6(a)). While room-temperature (300 K) simulations may predict similar initial melt depths, the resulting phase is mainly crystalline rather than amorphous as seen in experiment. This divergence underscores that amorphization is dictated by the cooling rate.

These cooling dynamics are also rooted in the temperature-dependent shift of ionization pathways. Cryogenic cooling modulates the initial energy coupling by suppressing the phonon-assisted single-photon absorption (SPA) pathways necessary for 1030 nm photons to bridge the indirect bandgap, while the two photon pathway is less affected by temperature. This reduction in phonon-mediated transitions forces carrier generation to be driven primarily by nonlinear two-photon absorption (TPA) at high intensities. The combination of this altered photon energy absorption and the significantly higher thermal conductivity of silicon at cryogenic temperatures yields a quench rate approximately 40% greater than at room temperature. This intensified cooling allows the molten volume to freeze into a stable amorphous state before epitaxial recrystallization can occur, enabling the formation of robust amorphous features.

## IV. CONCLUSIONS

In this work, cryogenic temperatures have been demonstrated to fundamentally alter the ultrafast laser-induced phase transition pathways in crystalline silicon. By extending experimental observations down to 24 K, a pronounced enhancement in amorphization, characterized by the emergence of a robust, bright annular ring surrounding the central damage site was identified. Analysis revealed a clear split in damage thresholds at lower temperatures, where the widening thermodynamic "window" between the onset of melting and the threshold for ablation allows the amorphous phase to stabilize in regions where recrystallization would typically dominate at room temperature. These findings are supported by spatially resolved Raman spectroscopy, which confirms that the amorphous layer thickness reaches a maximum of ~35 nm at the lowest temperatures, indicating that the extreme cold outpaces the naturally fast recrystallization velocity of the Si (100) orientation.

The physical mechanisms driving this behavior were discovered through a carrier-dependent two-temperature model (nTTM), which captures the transition from phonon-assisted linear absorption to nonlinear ionization pathways and changes in thermal transport at cryogenic conditions. Simulations show that the significantly enhanced thermal conductivity $K_l$ at 30 K drives a quench rate approximately 40% higher than at room temperature, facilitating an ultrafast extraction of heat that pushes the solidification front velocity beyond the critical threshold required for amorphous freezing. This study provides a comprehensive framework for understanding how temperature-governed mechanisms, such as phonon freeze-out and modified lattice relaxation dynamics, can be leveraged to achieve precision material modification. Beyond fundamental physics, these results suggest that cryogenic laser processing could serve as a powerful tool for manufacturing high-quality, buried photonic structures or for applications in extreme environments, such as space-based manufacturing.

## ACKNOWLEDGMENTS

AFOSR FA9550-25-1-0297 and The Ohio State University Presidential Catalyst Research funds for providing partial support. Electron microscopy was performed by Daniel Veghte at the Center for Electron Microscopy and Analysis (CEMAS) at The Ohio State University.